\documentclass{IEEEtran}
\usepackage{cite}
\usepackage{amsmath,amssymb,amsfonts}

\usepackage{graphicx}
\usepackage{textcomp}
\usepackage{amsmath,amsfonts}
\usepackage{algorithm}
\usepackage{algpseudocode}
\usepackage{tabularx}
\usepackage{subfigure}
\usepackage{subcaption}
\usepackage{booktabs}
\usepackage{siunitx}
\usepackage{placeins}
\usepackage{makecell}
\usepackage{amsthm}
\usepackage{setspace}
\def\BibTeX{{\rmB\kern-.05em{\sci\kern-.025emb}\kern-.08emT\kern-.1667em\lower.7ex\hbox{E}\kern-.125emX}}
    
\begin{document}
\title{Anti-Interference Communication Using Computational Antenna}
\author{Xiaocun Zong, \IEEEmembership{Student Member, IEEE}, Fan Yang*, \IEEEmembership{Fellow, IEEE}, \\Shenheng Xu, \IEEEmembership{Member, IEEE}, Maokun Li, \IEEEmembership{Fellow, IEEE}
\thanks{This work is supported by the National Key Research and Development
Program of China under Grant No. 2023YFB3811501. (Corresponding Author: \textit{Fan Yang})}
\thanks{
The authors are with the Department of Electronic Engineering, State Key laboratory of Space Network and Communications, Tsinghua University, Beijing 100084, China.
}

}
\maketitle

\begin{abstract}
This paper proposes a novel anti-interference communication method leveraging computational antennas, utilizing time averaging and 1-bit reconfigurable intelligent surfaces (RIS) to achieve robust signal modulation with minimal hardware complexity. We develop a communication model for computational antennas and propose an efficient signal processing algorithm optimized for temporal modulation. A USRP-based experimental platform is established to validate the approach under strong interference conditions (e.g., 5 dB jamming-to-signal ratio). Experimental results reveal up to an 80.9\% reduction in bit error rate (BER) and effective restoration of distorted images in transmission tests. Compared to conventional techniques like spread spectrum or frequency hopping, which require significant spectral resources, our method offers superior anti-interference performance without additional spectral overhead. This research provides valuable insights for radar detection, military communications, and next-generation wireless networks.
\end{abstract}


\section{Introduction}
\IEEEPARstart{T}{he} rapid proliferation of wireless devices and the escalating complexity of electromagnetic environments have significantly heightened the need for robust anti-interference techniques in modern wireless communication systems. Applications such as radar detection, military communications, and emerging sixth-generation (6G) wireless networks demand reliable signal transmission amidst challenging interference conditions \cite{b1,b2}. Conventional anti-interference methods, including spread spectrum, frequency hopping, often require additional spectral resources or sophisticated hardware, leading to increased system overhead, reduced spectral efficiency, and higher implementation costs \cite{b3,b4}.

Recent advancements in reconfigurable intelligent surfaces (RIS) have demonstrated transformative potential for enhancing wireless communication by enabling dynamic control over the propagation environment \cite{b10,b11,b12,b13}. By manipulating the phase and amplitude of reflected signals, RIS can optimize signal paths, improve coverage, and mitigate interference. However, their application in anti-interference scenarios is constrained by the requirement for multi-bit phase control, which introduces significant computational complexity and hardware costs \cite{b5,b6}. Similarly, time-modulated array antennas (TMAAs) have been explored for their ability to dynamically shape radiation patterns through temporal modulation. Despite their flexibility, TMAAs often prioritize harmonic frequencies, leaving the fundamental frequency underutilized for anti-interference purposes, thus limiting their effectiveness in complex electromagnetic environments \cite{b7,b8,b9}.

To overcome these challenges, we propose a novel paradigm termed \textit{Computational Antennas}, which leverages precise temporal control through a 1-bit RIS to optimize radiation characteristics at the fundamental frequency. By employing a time-averaging technique with carefully designed phase constants, our approach significantly reduces equivalent sidelobe levels, enhancing antenna directivity and system efficiency without requiring additional spectral resources. This computationally efficient method offers a breakthrough in achieving robust anti-interference performance, making it well-suited for dynamic and interference-heavy environments.

In this paper, we present a comprehensive theoretical and experimental investigation into the application of Computational Antennas for anti-interference communication. Our contributions are threefold: First, we develop a novel communication model for Computational Antennas, integrating a signal processing algorithm optimized for temporal modulation to enhance the signal-to-noise ratio (SNR) or signal-to-interference-plus-noise ratio (SINR). Second, we establish a Universal Software Radio Peripheral (USRP)-based experimental platform to validate the proposed method under strong interference conditions, with jamming-to-signal ratios (JSR) of 0 dB and 5 dB. Third, we demonstrate that our approach achieves up to an 80.9\% reduction in bit error rate (BER) and effectively restores distorted images in transmission experiments, surpassing the performance of conventional anti-interference techniques. These findings provide significant insights for applications in radar detection, military communications, and next-generation wireless systems, paving the way for more resilient and efficient communication technologies.

\section{Modeling and Theoretical Analysis}
\subsection{The Principle of Computational Antennas}
For a reconfigurable intelligent surface comprising \( M \times N \) elements, the array radiation pattern is expressed as (1), where the parameter \( A_{mn} \) represents the amplitude of the \( (m,n) \)-th element, \( \phi_{mn}^{ini} \) denotes the initial phase, \( \phi_{mn}^{out} \) is the output phase, and \( \phi_{mn}^{Q} \) is the quantization phase of the \( (m,n) \)-th element.

\begin{equation}
E(\theta,\phi) = \sum_{m=1}^{M}\sum_{n=1}^{N} A_{mn} e^{j(\varphi_{mn}^{ini} + \varphi_{mn}^{out} + \varphi_{mn}^{Q})}
\end{equation}

In practical applications of 1-bit reconfigurable reflective arrays, a phase offset \( \Delta\varphi \) is introduced prior to quantization, influencing the post-quantization phase distribution. For antenna pattern calculations, four phase constants are defined: \( \Delta\varphi_1 = 0 \), \( \Delta\varphi_2 = \pi/4 \), \( \Delta\varphi_3 = \pi/2 \), and \( \Delta\varphi_4 = 3\pi/4 \). The corresponding radiation patterns \( E_1 \), \( E_2 \), \( E_3 \), and \( E_4 \) are obtained for these \( \Delta\varphi \) values. The average radiation pattern \( E_{ave} \) is then computed through time-averaging as:

\begin{equation}
E_{ave}(\theta,\phi) = \frac{1}{4}\left(E_1 + e^{-j\pi/4}E_2 + e^{-j\pi/2}E_3 + e^{-j3\pi/4}E_4\right)
\end{equation}

\begin{figure}[t]
	\centering
	\includegraphics[width=0.99\linewidth]{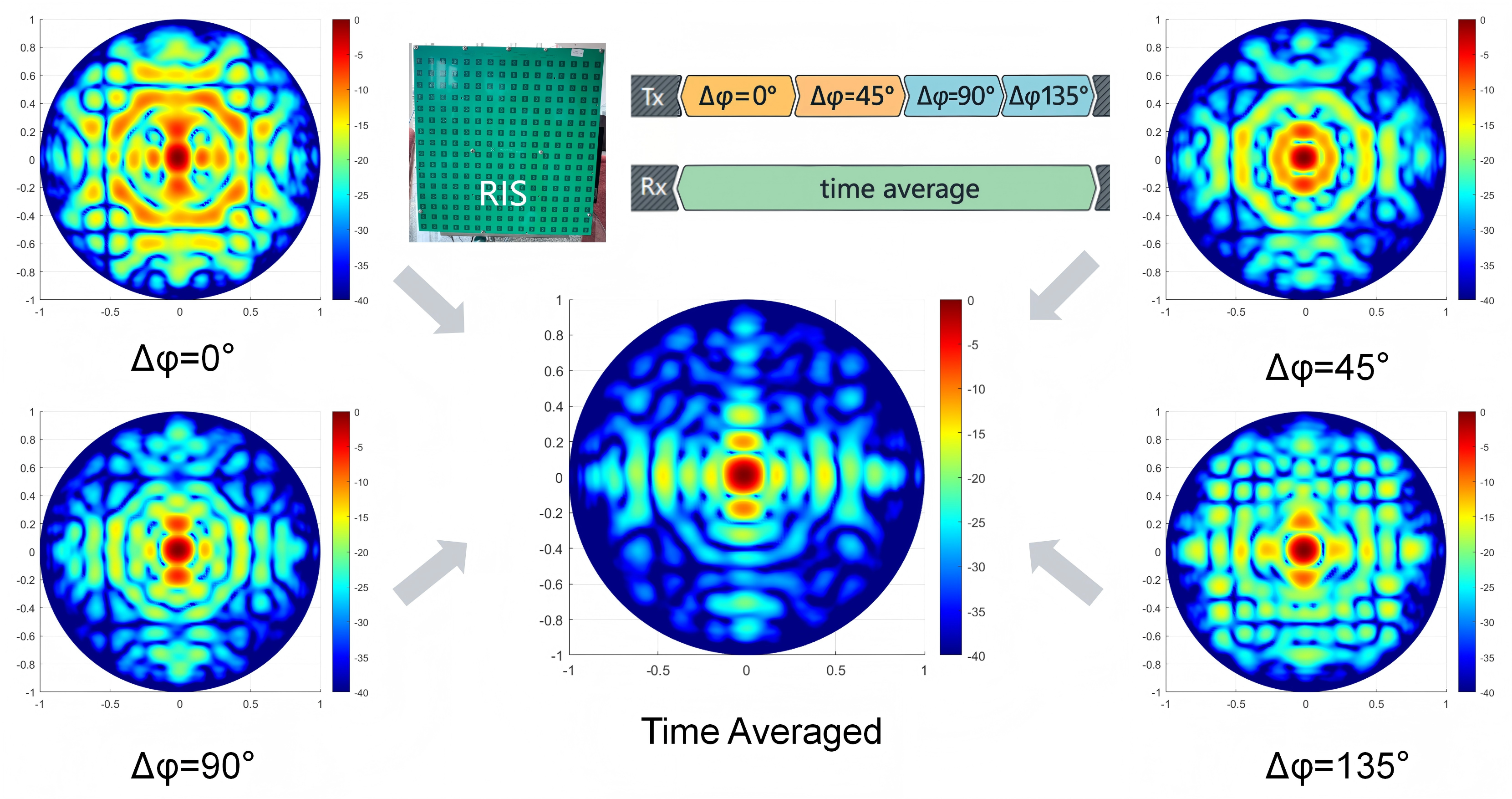}
	\caption{Time-averaged signal transmission and reception process diagram and the suppressed sidelobes.}
	\label{fig:tx-rx}
\end{figure}

Fig. \ref{fig:tx-rx} describes the principle of Computational Antennas. By employing the time-averaging method, the equivalent sidelobe levels of the radiation pattern can be effectively reduced.

\subsection{Communication Modeling}
Drawing on the concept of equivalent sidelobe reduction via computational antennas, we propose a communication model based on computational antennas. The transmitting antenna employs a 1-bit RIS and transmits four identical data streams using the previously described phase constant control method. At the receiving end, a dedicated algorithm processes the received signal to enhance the equivalent SNR or signal-to-interference-plus-noise ratio (SINR).

\begin{figure*}[t]
	\centering
	\includegraphics[width=0.8\linewidth]{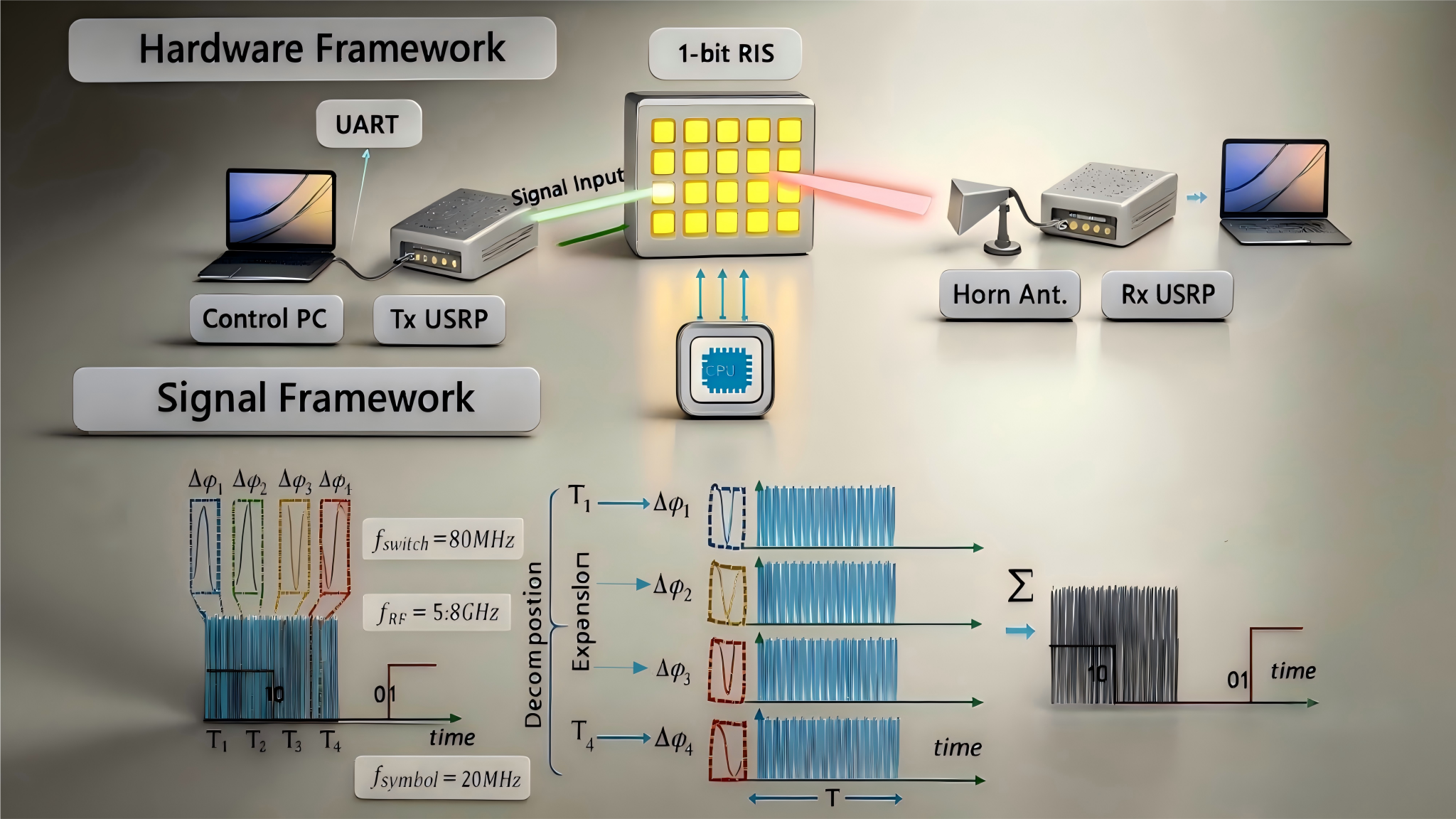}
	\caption{Framework of the Computational Antennas communication system.}
	\label{fig:framework}
\end{figure*}

For instance, in a continuous communication scenario in Fig. \ref{fig:framework} with a carrier frequency of 5.8 GHz, a symbol rate of 20 MHz, and an 80 MHz phase constant switching rate (The phase constant switching rate must be four times the symbol rate to enable each symbol to be divided into four equal parts, followed by time-stretching and weighted averaging.), each symbol is divided into four segments for transmission. At the receiver, the symbol is segmented into four equal parts, time-stretched, and weighted-averaged to reconstruct the equivalent communication signal within the computational antenna framework. To streamline experiments, real-time signal processing is conducted offline by transmitting the same signal four times with different phase constants, followed by post-processing of the resulting signals. 

The aforementioned transmission and reception processes are reciprocal, and the reconfigurable intelligent surface can achieve equivalent performance when employed for reception.

\section{Signal Processing Algorithm}

\begin{algorithm}[t]
    \centering
    \begin{minipage}{0.99\linewidth}
        \caption{Advanced Maximum Ratio Combining (MRC) for Computational Antenna Multi-Segment QPSK Signal Fusion and Robust Enhancement}
        \label{alg:advanced_mrc_qpsk}
        \renewcommand{\algorithmicrequire}{\textbf{Input:}}
        \renewcommand{\algorithmicensure}{\textbf{Output:}}
        \begin{algorithmic}[1]
            \Require Four received QPSK signal segments $\{r_k[n]\}_{k=0}^3 \in \mathbb{C}^{N}$, segment-specific signal quality metrics $\{\textit{SNR}_k\}_{k=0}^3 \in \mathbb{R}^+$ (or $\textit{SINR}_k$ in interference-dominated regimes)
            \Ensure Enhanced Computational Antenna QPSK signal $r_{\text{merged}}[n] \in \mathbb{C}^{N}$, yielding maximal post-combination SNR/SINR via adaptive diversity exploitation
           
            \State Identify the reference segment $r_0[n]$ as the one with the supremum SNR: $r_0[n] \leftarrow \arg\max_{k \in \{0,1,2,3\}} \textit{SNR}_k$
          
            \Comment{Selection of the highest-SNR segment as the intrinsic reference ensures alignment fidelity without external priors, leveraging inherent channel diversity.}
           
            \State Initialize the fused signal accumulator: $r_{\text{merged}}[n] \leftarrow \mathbf{0}_N$ for all $n = 1, \dots, N$
           
            \For{$k = 0$ \textbf{to} $3$}
                \State Compute the phase offset $\theta_k$ as the argument of the complex cross-correlation between $r_k[n]$ and the reference $r_0[n]$:
                \[
                \theta_k = \angle \left( \sum_{n=1}^N r_k[n] \, r_0^*[n] \right)
                \]
               
               
                \State Derive the normalized amplitude weight $w_k$ as the fractional contribution of the $k$-th segment's SNR to the aggregate:
                \[
                w_k = \frac{\textit{SNR}_k}{\sum_{i=0}^3 \textit{SNR}_i}
                \]
                This proportionality scheme, rooted in MRC optimality, amplifies reliable segments while mitigating the impact of degraded ones, with $\sum_k w_k = 1$.
               
                \State Generate the phase-compensated and weighted segment: $r_k'[n] \leftarrow w_k \, r_k[n] \, e^{j \theta_k}$ for all $n = 1, \dots, N$
               
                \State Incrementally fuse the compensated segment: $r_{\text{merged}}[n] \leftarrow r_{\text{merged}}[n] + r_k'[n]$ for all $n = 1, \dots, N$
            \EndFor
           
            \State \Return $r_{\text{merged}}[n]$, the coherence-enhanced computational antenna signal primed for downstream synchronization and demodulation
        \end{algorithmic}
    \end{minipage}
\end{algorithm}
The time-averaging method, introduces distinct phase constants to the antenna’s radiation patterns. In the context of communication systems, this approach results in signals with varying initial phases across four time segments, each corresponding to a phase constant (\(\Delta\varphi = 0^\circ, 45^\circ, 90^\circ, 135^\circ\)). To mitigate the impact of these phase variations, the receiver must align the signals before combining them. Unlike antenna measurements, where signals with different phase constants have uniform amplitude, real-world communication scenarios exhibit varying signal strengths due to channel effects and noise. Consequently, direct summation of these signals is suboptimal; instead, we employ maximum ratio combining (MRC) to perform weighted averaging.

The MRC process involves computing complex weights for each of the four signal segments. The phase of each weight, \(\theta_k\), is determined by the cross-correlation between the received signal segment \(r_k[n]\) and a reference signal \(r_0[n]\), as shown in Equation (3). The amplitude of each weight, \(w_k\), is calculated as the proportion of the SNR of the \(k\)-th segment to the total SNR across all segments, as given in Equation (3). The merged signal, \(r_{merged}\), is then obtained by weighted averaging, as expressed in Equation (\ref{eq:r_merged}). This process, which can be seen in Algorithm \ref{alg:advanced_mrc_qpsk}, effectively enhances the signal quality by constructively combining the in-phase components while suppressing noise or interference through anti-phase cancellation. In the presence of interference during the communication process, the signal-to-noise ratio (SNR) is replaced by the signal-to-interference-plus-noise ratio (SINR). Essentially, this is a weighted averaging method based on signal strength.

\begin{equation}
    \theta_k = \angle \left( \sum_{n=1}^N r_k[n] r_0^*[n] \label{eq:theta_k}\right),\ \  w_k = \frac{\textit{SNR}_k}{\sum_{i=0}^3 \textit{SNR}_i} 
\end{equation}

\begin{equation}
    r_{merged} = \sum_{k=0}^3 r_k[n] \cdot w_k e^{j \theta_k} \label{eq:r_merged}
\end{equation}

For online continuous transmission processing, the signal processing at the receiving end involves dividing each symbol into four equal parts, replicating each symbol four times its original length, and then performing weighted averaging of the four signals to maintain consistent signal length. In offline local processing, the same signal is transmitted four times, and the four received signals are processed locally through weighted averaging.

\section{Experimental Validation}
\begin{figure}[t]
	\centering
	\includegraphics[width=0.95\linewidth]{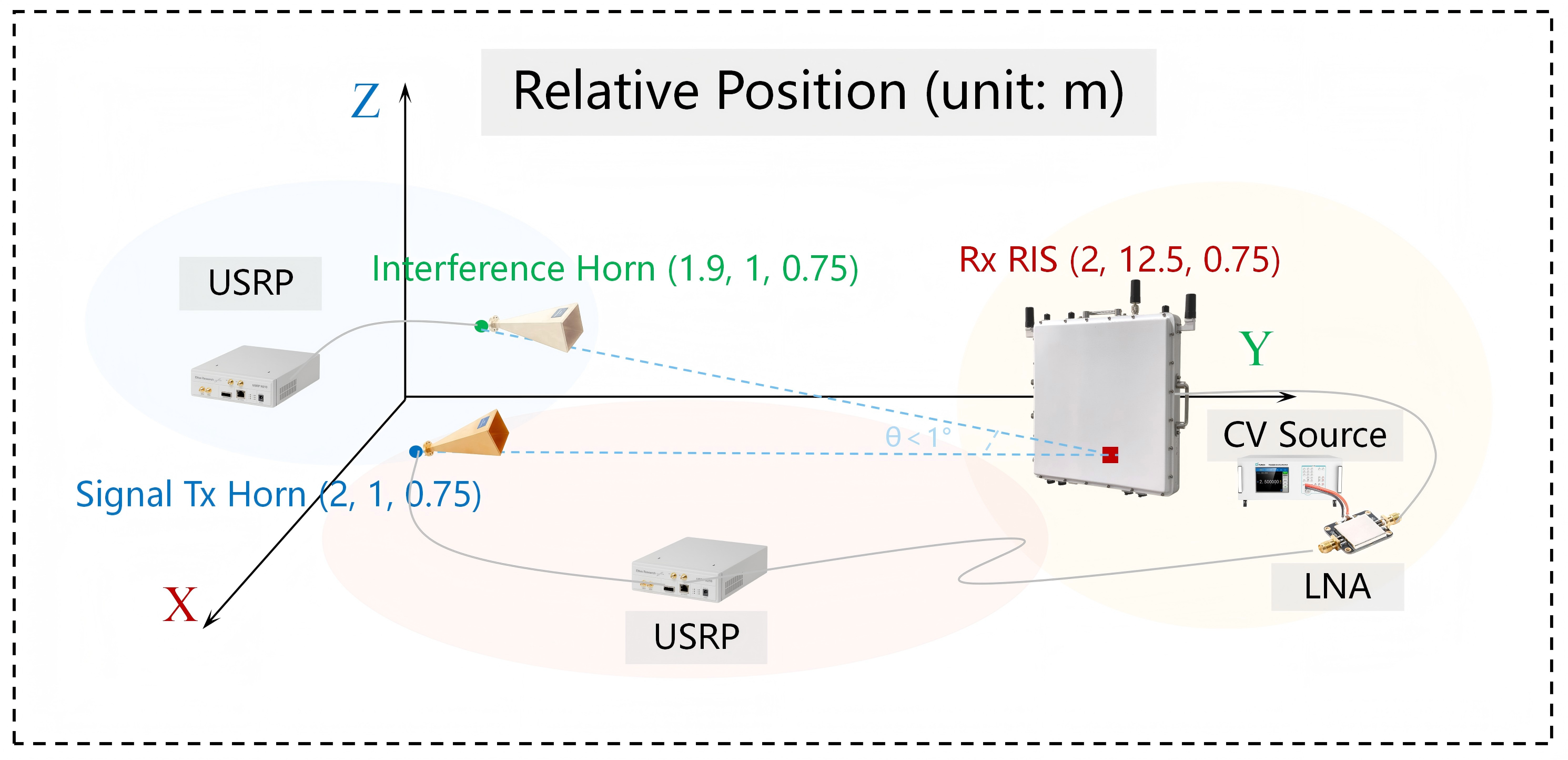}
	\caption{Relative positions of the signal source, interference source, and RIS in the experimental setup.}
	\label{fig:position}
\end{figure}

An experimental platform based on a Universal Software Radio Peripheral (USRP) was developed to validate the proposed method. The setup comprises a Computational Antenna Reconfigurable Intelligent Surface (RIS) with a $16\times 16$ element array operating at 5.8GHz, paired with a USRP-Y230 software-defined radio. The experimental configuration, illustrated in Fig.~\ref{fig:position}, depicts the relative positions of the signal source, interference source, and RIS. The signal and interference sources are co-located, emitting signals at the same frequency and direction. To compensate for the low transmit power of the USRP, a low-noise amplifier is integrated at the receiver to enhance signal strength.

\begin{figure}[t]
	\centering
	\includegraphics[width=0.95\linewidth]{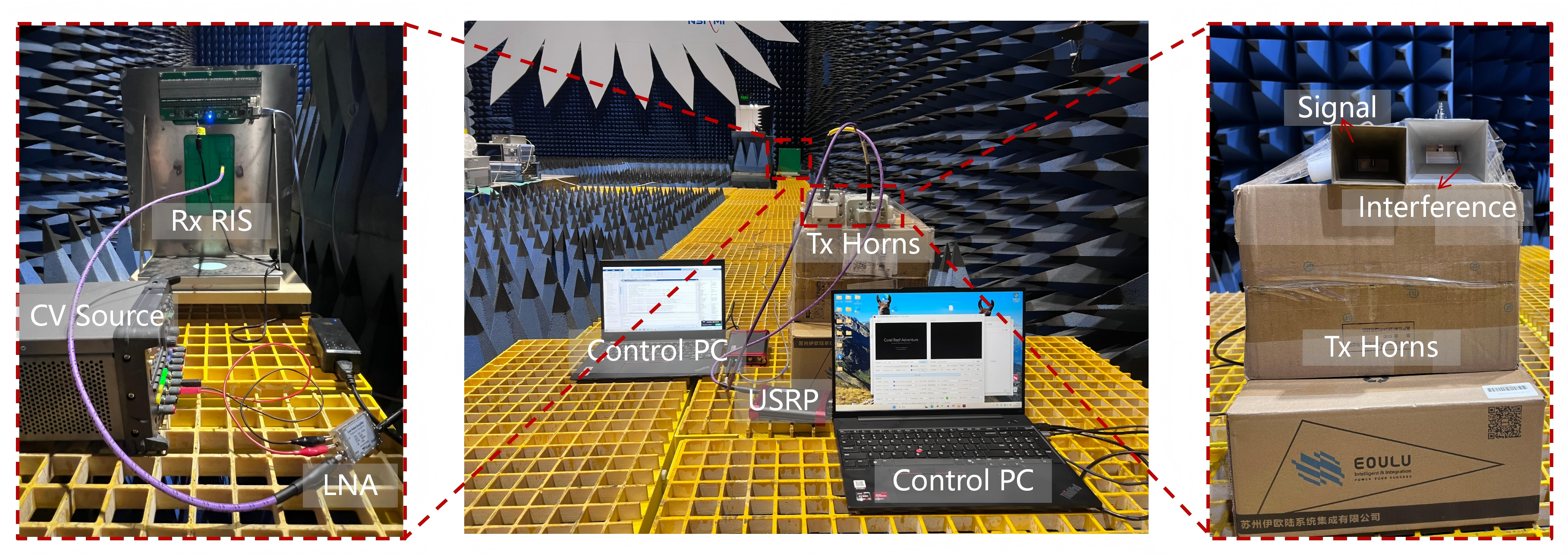}
	\caption{Experimental environment in the B3 anechoic chamber.}
	\label{fig:environment}
\end{figure}

The experiments were conducted in the B3 anechoic chamber at the Rohm Building, Tsinghua University, to ensure a controlled environment with high signal fidelity. The final configuration of the experimental platform is shown in Fig.~\ref{fig:environment}.

Experiments were performed under three conditions: no interference, equal interference and signal strength (Jamming-to-Signal Ratio, JSR = \SI{0}{\decibel}), and stronger interference (JSR = \SI{5}{\decibel}). These tests aimed to verify improvements in signal-to-noise ratio (SNR) under no interference, signal-to-interference ratio under weak interference, and the effectiveness of interference cancellation under strong interference. The experimental parameters included a center frequency of \SI{5.8}{\giga\hertz}, a horn antenna gain of \SI{15}{\decibel}, and a transmission power of \SI{-20} dBm. The results are summarized in the TABLE \uppercase\expandafter{\romannumeral1}.
\begin{figure*}[t]
	\centering
	\includegraphics[width=0.98\linewidth]{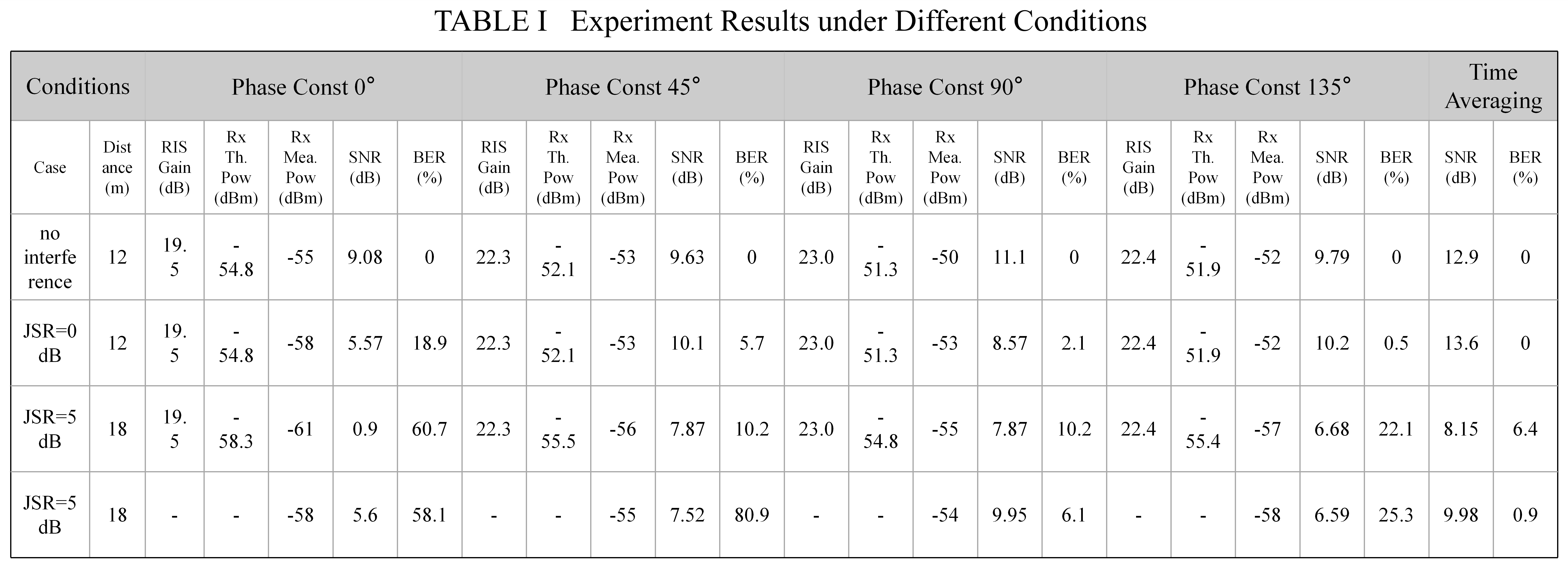}
	\label{fig:table}
\end{figure*}

\begin{figure}[t]
	\centering
	\includegraphics[width=0.95\linewidth]{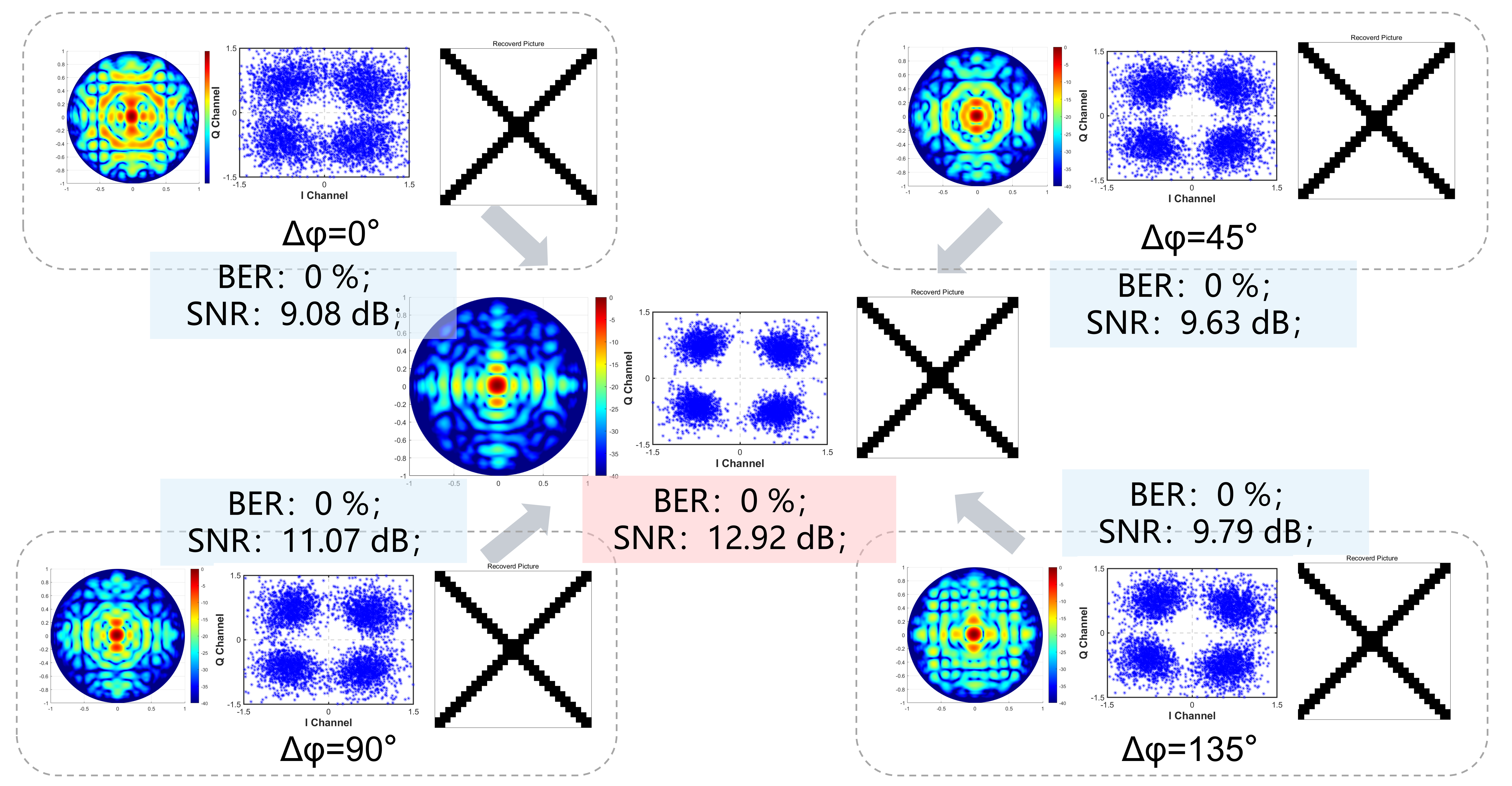}
	\caption{Experimental results under no interference, demonstrating high-quality signal transmission.}
	\label{fig:no_interferece}
\end{figure}

As shown in Fig.~\ref{fig:no_interferece}, in the absence of interference, the signal quality was excellent, enabling perfect image transmission. The computational antenna algorithm improved the SNR by \SIrange{1.85}{3.84}{\decibel}.

\begin{figure}[t]
	\centering
	\includegraphics[width=0.95\linewidth]{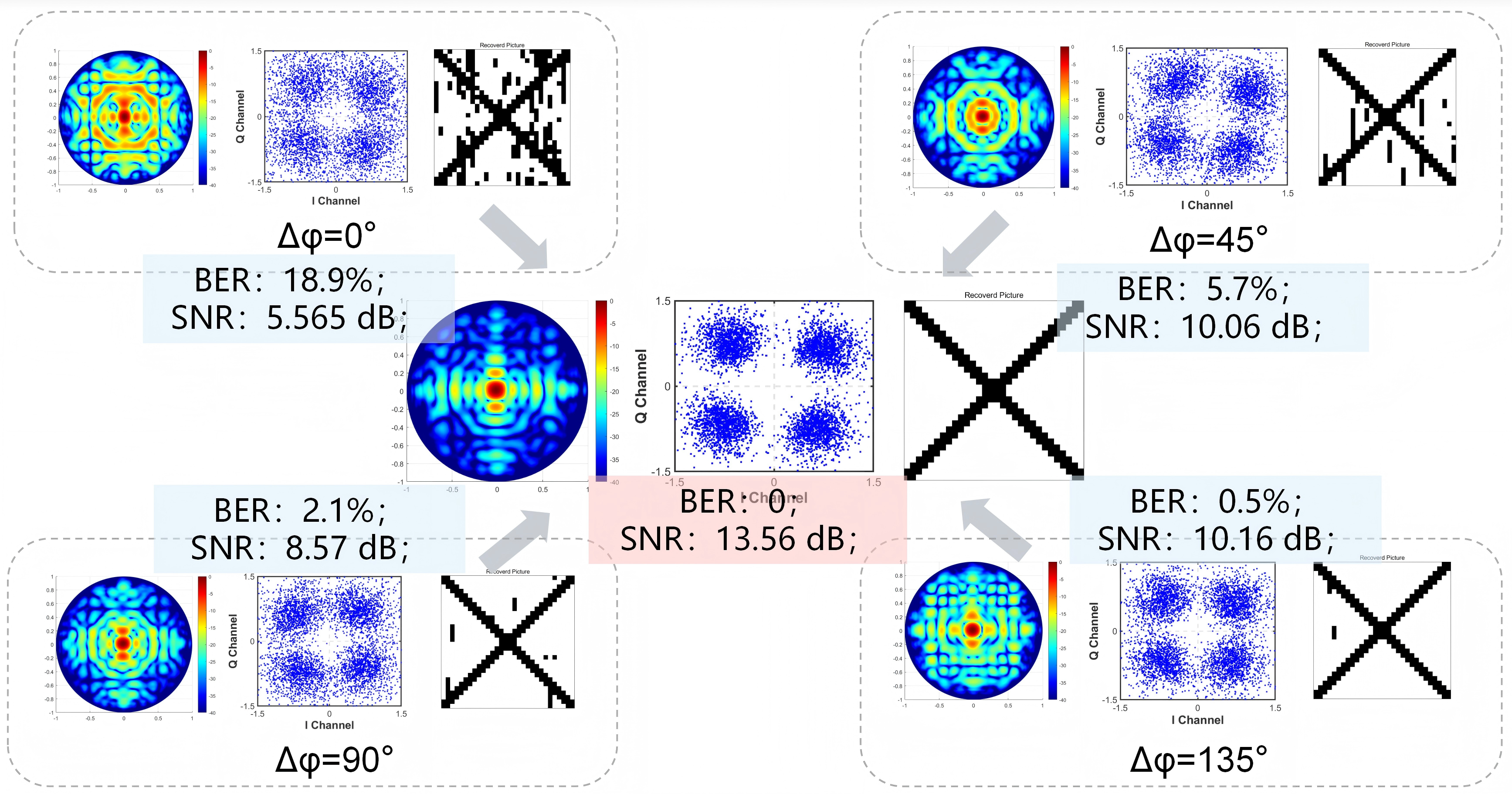}
	\caption{Experimental results at JSR = \SI{0}{\decibel}, showing restored image quality after signal processing.}
	\label{fig:JSR0dB}
\end{figure}

Fig.~\ref{fig:JSR0dB} illustrates the results when the interference and signal strengths are equal (JSR = \SI{0}{\decibel}). Minor degradation in image quality was observed, leading to slight bit errors. Performance was evaluated through the antenna radiation pattern, signal reception constellation diagram, and image quality across various phase constants with time-averaging applied. Post-processing with the computational antenna improved the SNR by \SIrange{3.4}{8}{\decibel}, reduced the bit error rate by up to \SI{18.9}{\percent}, and fully restored the distorted image, confirming the method's efficacy in mitigating interference and enhancing communication reliability.

\begin{figure}[t]
	\centering
	\includegraphics[width=0.95\linewidth]{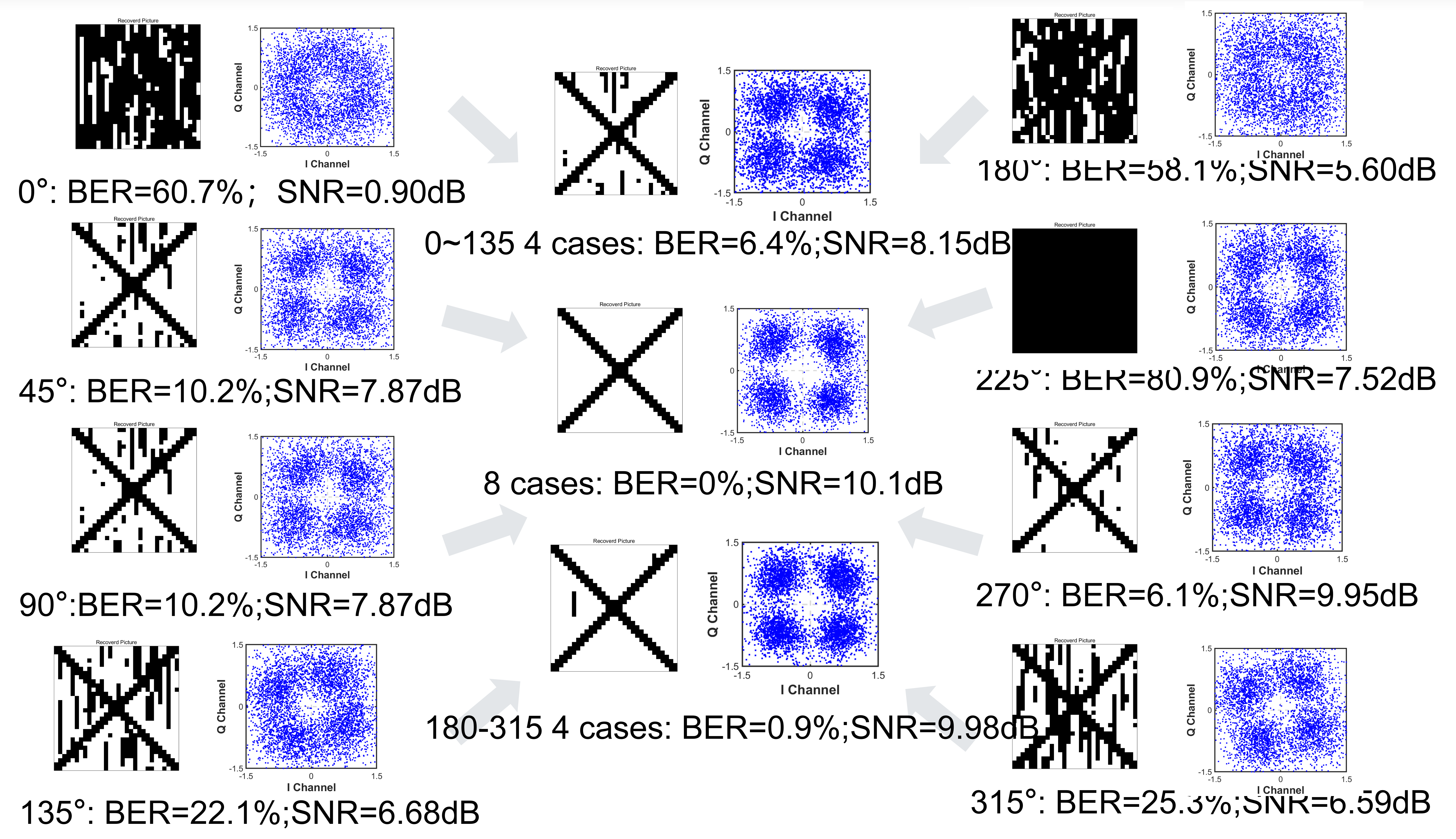}
	\caption{Experimental results at JSR = \SI{5}{\decibel}, showing complete image restoration after enhanced signal processing.}
	\label{fig:JSR5dB}
\end{figure}

When the JSR increased to \SI{5}{\decibel}, as depicted in Fig.~\ref{fig:JSR5dB}, channel quality deteriorated significantly due to intensified interference, resulting in complete image distortion. With four phase constant configurations, time-averaging reduced the bit error rate to some extent, though minor errors persisted. By introducing four additional phase constant experiments, the computational antenna processing reduced the bit error rate by up to \SI{80.9}{\percent}, fully restoring the distorted image and demonstrating the robustness of the proposed method under strong interference conditions. The reason why the introduction of eight sets of phase constants can improve the anti-interference performance is that the interference phases differ by 180° between each other, achieving phase cancellation, thereby improving the signal-to-interference ratio.

Under interference conditions, the most significant improvement was observed in the signal-to-interference-plus-noise ratio (SINR). Although precise SINR data could not be obtained due to limitations in the current computational software, the successful image recovery clearly indicates a substantial enhancement in SINR, underscoring the effectiveness of the proposed method.

\section{Analysis and Discussions}
\begin{figure}[t]
	\centering
	\includegraphics[width=0.9\linewidth]{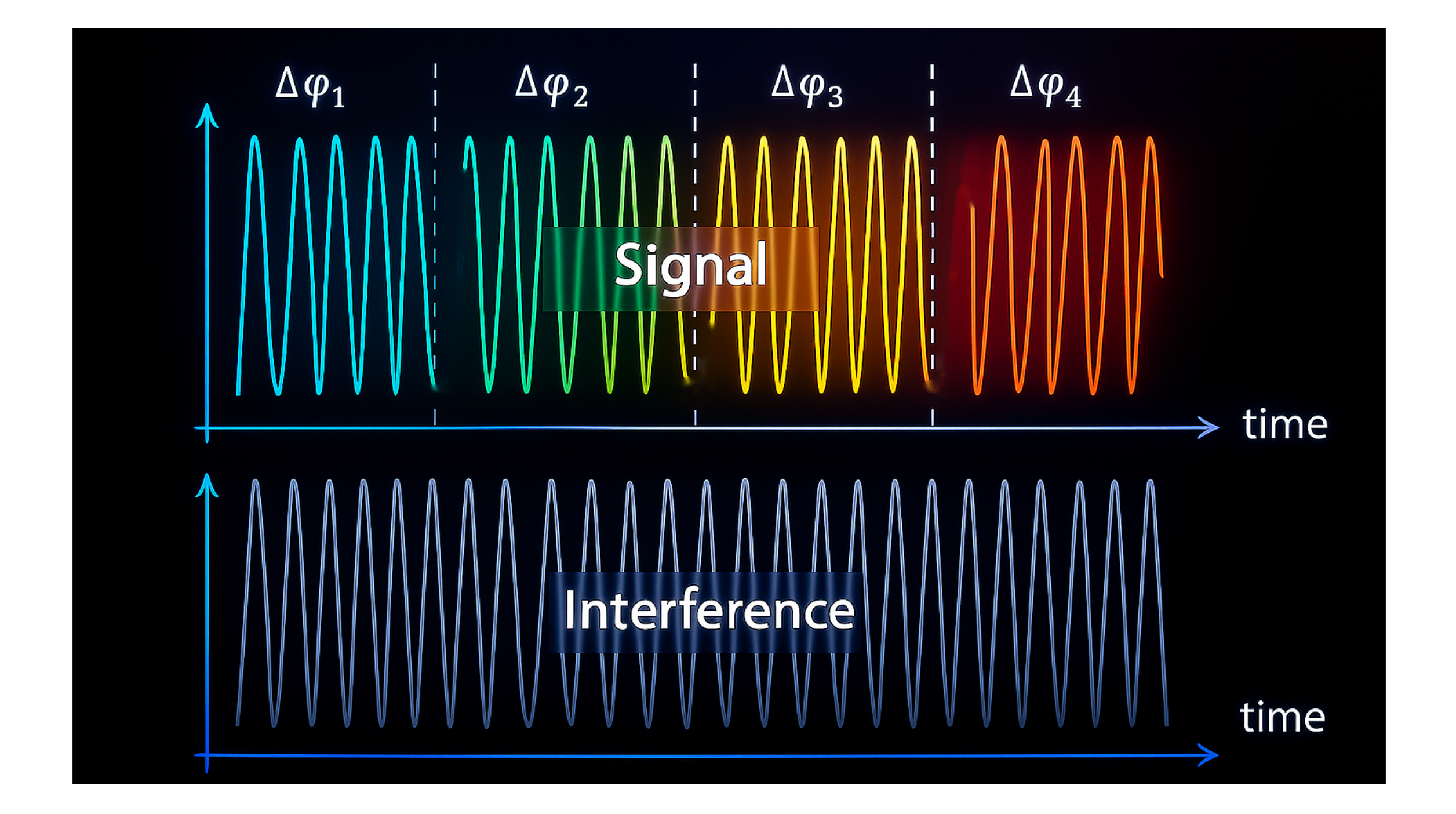}
	\caption{Contrast between computational antenna signal and interference.}
	\label{fig:contrast}
\end{figure}

To elucidate the distinctions between conventional communication signals and those modulated by computational antennas, employing the same signal processing algorithm, the following points are analyzed:
\begin{itemize}
    \item In the absence of interference, weighted averaging of the four signals results in both the conventional signal and the computational antenna-modulated signal being in phase. However, as noise is random, it remains statistically indistinguishable after averaging. Consequently, the signal-to-noise ratio (SNR) for both signal types is enhanced.
    \item In the presence of interference, as shown in Fig.~\ref{fig:contrast}, the continuous interference signal maintains a consistent initial phase across its four equally divided segments. In contrast, the four communication signals modulated by computational antennas possess distinct initial phases. After alignment, the original signals become in-phase, while the interference signal, due to its uniform initial phase, undergoes phase shifts that lead to mutual cancellation. This process significantly improves the signal-to-interference-plus-noise ratio (SINR) and overall signal quality. However, for conventional signals, both the signal and interference are phase-continuous. As a result, after dividing into four equal parts and aligning, the signal and interference become in-phase and constructively combine, leading to no improvement in the signal-to-interference-plus-noise ratio.
\end{itemize}

Thus, our proposed algorithm, leveraging computational antennas, enhances the SNR under non-interference conditions and the SIR in the presence of interference. Although it provides no anti-interference advantages for conventional signals, it still improves SNR to a certain extent. This underscores the versatility and robustness of our algorithm across diverse signal environments. A rigorous mathematical proof of the SNR and SIR improvement is attached in the Appendix~\ref{app:snr_proof} and~\ref{app:sir_proof}  .

\section{Conclusions}
This paper introduces a novel anti-interference communication approach utilizing Computational Antennas, which employs time-averaging with a 1-bit Reconfigurable Intelligent Surface (RIS) to enhance signal robustness without additional spectral resources. Through rigorous theoretical modeling and experimental validation on a USRP-based platform under strong interference conditions (JSR of 0 dB and 5 dB), the method achieves up to an 80.9\% reduction in bit error rate (BER) and effectively restores distorted images in transmission tests. These outcomes demonstrate the method’s capability to maintain reliable communication in challenging electromagnetic environments. The findings position Computational Antennas as a promising solution for applications in radar detection, military communications, and future wireless systems like 6G. Future research will aim to enhance the communication protocol to enable real-time online processing of signal algorithms, improve computational efficiency and investigate the method’s adaptability across a broader range of operational scenarios.


\appendices

\section{Proof of SNR Improvement}
\label{app:snr_proof}

Repeating a QPSK symbol four times with distinct phase offsets, followed by phase alignment and MRC, increases the SNR by a factor of 4 in an AWGN channel.

\begin{proof}
Consider a QPSK symbol \( s \in \mathbb{C} \) with power \( P_s = |s|^2 \), transmitted four times over an AWGN channel. The \( k \)-th transmission (\( k = 1, 2, 3, 4 \)) has phase offset \( \theta_k \), so the transmitted signal is \( s e^{j \theta_k} \). The received signal is:
\begin{equation}
    r_k = s e^{j \theta_k} + n_k,
\end{equation}
where \( n_k \sim \mathcal{CN}(0, \sigma^2) \) is independent complex Gaussian noise with power \( \sigma^2 \). The single-transmission SNR is:
\begin{equation}
    \text{SNR}_1 = \frac{P_s}{\sigma^2} = \frac{|s|^2}{\sigma^2}.
\end{equation}

At the receiver, phase alignment multiplies \( r_k \) by \( e^{-j \theta_k} \):
\begin{equation}
    r_k' = r_k e^{-j \theta_k} = s + n_k e^{-j \theta_k} = s + n_k',
\end{equation}
where \( n_k' \sim \mathcal{CN}(0, \sigma^2) \), as the noise distribution is rotationally invariant.

MRC in AWGN simplifies to equal-gain combining. The combined signal is:
\begin{equation}
    r_{\text{MRC}} = \frac{1}{4} \sum_{k=1}^4 r_k' = s + \frac{1}{4} \sum_{k=1}^4 n_k'.
\end{equation}
- \textbf{Signal component}: \( s \), power \( P_s = |s|^2 \).
- \textbf{Noise component}: \( n_{\text{MRC}} = \frac{1}{4} \sum_{k=1}^4 n_k' \). Since \( n_k' \) are independent and \( \mathcal{CN}(0, \sigma^2) \),
\begin{equation}
    \text{Var}(n_{\text{MRC}}) = \text{Var}\left( \frac{1}{4} \sum_{k=1}^4 n_k' \right) = \frac{1}{16} \cdot 4 \sigma^2 = \frac{\sigma^2}{4}.
\end{equation}

The combined SNR is:
\begin{align}
    \text{SNR}_{\text{MRC}} &= \frac{P_s}{\text{Var}(n_{\text{MRC}})} = \frac{|s|^2}{\sigma^2 / 4} \notag \\
    &= 4 \cdot \frac{|s|^2}{\sigma^2} = 4 \cdot \text{SNR}_1.
\end{align}
Thus, the SNR improves by a factor of 4 (6 dB in logarithmic scale).

The improvement arises because phase alignment ensures coherent signal addition (preserving \( P_s \)), while noise adds incoherently, reducing its power by a factor of 4 due to averaging.
\end{proof}


\section{Theoretical Proof of Signal-to-Interference Ratio Improvement}
\label{app:sir_proof}
\subsection{System Model}
Consider a QPSK symbol \( s \in \mathbb{C} \), power \( P_s = |s|^2 \), transmitted four times over an AWGN channel with distinct phase offsets \( \theta_k \) (\( k = 1, 2, 3, 4 \)), yielding transmitted signal \( s e^{j \theta_k} \). The channel includes:
\begin{itemize}
    \item \textbf{Interference}: Continuous narrowband signal \( i_k = A e^{j \phi_k} \), where \( \phi_k = \phi_1 + (k-1) \Delta \phi \), \( \Delta \phi = \omega T \), and interference power \( P_i = |A|^2 \).
    \item \textbf{Noise}: \( n_k \sim \mathcal{CN}(0, \sigma^2) \), independent.
\end{itemize}
Received signal:
\begin{equation}
    r_k = s e^{j \theta_k} + i_k + n_k.
\end{equation}
Single-transmission SIR:
\begin{equation}
    \text{SIR}_1 = \frac{P_s}{P_i} = \frac{|s|^2}{|A|^2}.
\end{equation}

\subsection{Receiver Processing}
Phase alignment multiplies \( r_k \) by \( e^{-j \theta_k} \):
\begin{equation}
    r_k' = s + i_k' + n_k', \quad i_k' = A e^{j (\phi_k - \theta_k)}, \quad n_k' \sim \mathcal{CN}(0, \sigma^2).
\end{equation}
MRC (equal-gain combining in AWGN) yields:
\begin{equation}
    r_{\text{MRC}} = \frac{1}{4} \sum_{k=1}^4 r_k' = s + \frac{1}{4} \sum_{k=1}^4 i_k' + \frac{1}{4} \sum_{k=1}^4 n_k'.
\end{equation}
Combined SIR:
\begin{equation}
    \text{SIR}_{\text{MRC}} = \frac{P_s}{\left| \frac{1}{4} \sum_{k=1}^4 i_k' \right|^2}.
\end{equation}

\subsection{Cauchy-Schwarz Inequality in Interference Power}

The SIR after fourfold repetition and MRC is at least four times the single-transmission SIR, with the Cauchy-Schwarz inequality providing the interference power upper bound.

\begin{proof}
The combined interference is:
\begin{equation}
    \bar{i} = \frac{1}{4} \sum_{k=1}^4 i_k' = \frac{A}{4} \sum_{k=1}^4 e^{j (\phi_k - \theta_k)} = \frac{A}{4} e^{j \phi_1} \sum_{k=1}^4 e^{j \psi_k},
\end{equation}
where \( \psi_k = (k-1) \Delta \phi - \theta_k + \theta_1 \). Interference power:
\begin{equation}
    P_{\bar{i}} = |\bar{i}|^2 = \frac{A^2}{16} \left| \sum_{k=1}^4 e^{j \psi_k} \right|^2.
\end{equation}

To bound \( \left| \sum_{k=1}^4 e^{j \psi_k} \right|^2 \), apply the Cauchy-Schwarz inequality to the complex sum:
\begin{align}
    \left| \sum_{k=1}^4 e^{j \psi_k} \right|^2 = \left| \sum_{k=1}^4 1 \cdot e^{j \psi_k} \right|^2 \notag 
    &\leq \left( \sum_{k=1}^4 |1|^2 \right) \left( \sum_{k=1}^4 |e^{j \psi_k}|^2 \right) \notag \\
    &= 4 \cdot 4 = 16,
\end{align}
since \( |e^{j \psi_k}| = 1 \). Thus:
\begin{equation}
    P_{\bar{i}} \leq \frac{A^2}{16} \cdot 16 = A^2 = P_i.
\end{equation}
This upper bound occurs when \( e^{j \psi_k} \) are co-phased (e.g., \( \psi_k = \psi \)), yielding no SIR improvement.

However, the continuous interference (\( \phi_k \approx \phi_1 \)) and designed \( \theta_k \) (e.g., orthogonal or random) reduce this term:
\begin{itemize}
    \item \textbf{Orthogonal Phases}: Set \( \theta_k = \theta_1 + (k-1) \pi / 2 \), \( \Delta \phi \approx 0 \):
    \[
    \sum_{k=1}^4 e^{j (\phi_1 - \theta_k)} = e^{j \phi_1} (1 - j - 1 + j) = 0.
    \]
    Then \( P_{\bar{i}} = 0 \), \( \text{SIR}_{\text{MRC}} \to \infty \).
    \item \textbf{Random Phases}: If \( \theta_k \sim U[0, 2\pi) \), independently:
    \[
    E[P_{\bar{i}}] = \frac{A^2}{16} \sum_{k=1}^4 \sum_{m=1}^4 E[e^{j (\psi_k - \psi_m)}] = \frac{A^2}{16} \cdot 4 = \frac{P_i}{4},
    \]
    since \( E[e^{j (\theta_m - \theta_k)}] = \delta_{km} \). Thus:
    \[
    \text{SIR}_{\text{MRC}} = \frac{P_s}{P_i / 4} = 4 \cdot \text{SIR}_1.
    \]
\end{itemize}
The Cauchy-Schwarz bound shows the worst-case (no suppression), while phase design ensures \( \left| \sum e^{j \psi_k} \right|^2 < 16 \), improving SIR.
\end{proof}

\subsection{Analysis}
The Cauchy-Schwarz inequality provides the maximum interference power when phases align. By introducing orthogonal or random \( \theta_k \), the interference adds non-coherently or destructively, reducing \( P_{\bar{i}} \), while the signal adds coherently, increasing SIR by at least a factor of 4.

\end{document}